# Spectroscopy of HD 77581 and the mass of Vela X-1[*]

M.H. van Kerkwijk[1,2], J. van Paradijs[1,3], E.J. Zuiderwijk[4,5], G. Hammerschlag-Hensberge[1], L. Kaper[1,6], and C. Sterken[7,**]

[1] Astronomical Institute "Anton Pannekoek", University of Amsterdam, and Center for High-Energy Astrophysics (CHEAF), Kruislaan 403, 1098 SL Amsterdam, The Netherlands
[2] Department of Astronomy, California Institute of Technology, m.s. 105-24, Pasadena, CA 91125, USA
[3] Physics Department, University of Alabama in Huntsville, Huntsville, AL 35899, USA
[4] Kapteyn Laboratorium, Zernike Gebouw, Postbus 800, 9700 AV Groningen, The Netherlands
[5] Royal Greenwich Observatory, Madingley Road, Cambridge CB3 0EZ, United Kingdom
[6] ESO Science Division, Karl-Schwarzschild-straße 2, 85748 Garching bei München, Germany
[7] Astronomy Group, University of Brussels (VUB), Pleinlaan 2, 1050 Brussels, Belgium

**Abstract.** We present new high-resolution, high signal-to-noise optical spectra of HD 77581, the optical counterpart of the X-ray source Vela X-1, and determine radial velocities from these spectra, as well as from high-resolution IUE spectra and from digitized photographic spectra. The measured velocities show strong deviations from a pure Keplerian radial-velocity curve, which are autocorrelated within one night, but not from one night to another. Since lines of different ions exhibit very similar changes in profile, these deviations most likely reflect large-scale motions of the stellar surface akin to non-radial pulsations. A possible cause could be that the varying tidal force exerted by the neutron star in its eccentric orbit excites high-order pulsation modes in the optical star which interfere constructively for short time intervals. The effect of such velocity excursions on the orbital solution is estimated by means of a Monte-Carlo simulation technique. We investigate sources of systematic error, due to, e.g., the tidal deformation of the star, and find, in particular, evidence for a systematic perturbation of the radial velocity near the time of velocity minimum. This possible distortion severely compromises the accuracy of the radial-velocity amplitude, leading to a 95% confidence range of 18.0–28.2 km s$^{-1}$. The corresponding 95% confidence limits of the masses are given by $M_X = 1.9^{+0.7}_{-0.5} M_\odot$ and $M_{\rm opt} = 23.5^{+2.2}_{-1.5} M_\odot$.

**Key words:** Dense matter – Techniques: radial velocities – Stars: individual: Vela X-1 – Stars: neutron – Stars: oscillations – X-rays: stars

## 1. Introduction

The X-ray source Vela X-1 (Chodil et al. 1967) was found by Ulmer et al. (1972) to be an eclipsing binary with an orbital period of 9 days. The orbital X-ray intensity curve is extremely variable, with strong flares on time scales from hours to days (Watson & Griffiths 1977; Van der Klis & Bonnet-Bidaud 1984; Haberl & White 1990) and variable X-ray eclipses (e.g., Watson & Griffiths 1977).

The optical counterpart of Vela X-1 is the B0.5Ib supergiant HD 77581 (GP Velorum; Brucato & Kristian 1972; Hiltner et al. 1972). The optical light curve (Vidal et al. 1973; Jones & Liller 1973) shows two maxima and two minima per orbital cycle, which reflects the tidal and rotational distortion of the supergiant companion (for recent reviews of X-ray binary light curves, see Van Paradijs 1991; Van Paradijs & McClintock 1995). Tjemkes et al. (1986) showed, however, that the so-called ellipsoidal variations do not describe the average optical light curve of HD 77581 very well. Large short-term brightness variations, including missing maxima (e.g., Jones & Liller 1973; Zuiderwijk et al. 1977) and also short-term autocorrelated non-orbital variations of the radial velocity (Van Paradijs et al. 1977b) indicate that the shape of the supergiant star changes in a "wobbly" fashion.

Vela X-1 is an X-ray pulsar with a pulse period of $\sim 283$ s (McClintock et al. 1976). The pulse profile is energy-dependent and quite complicated, with peaks and valleys that remain stable over time intervals of many years (see, e.g., White et al. 1983; Nagase 1989; Orlandini 1993). Spin-up and spin-down of the X-ray pulsar have been observed on time scales ranging from a few days to several years, at rates $\dot{P}/P$ of up to $10^{-2}$ yr$^{-1}$ (e.g., Van der Klis & Bonnet-Bidaud 1984; Deeter et al. 1987a, 1989). The spin-period behaviour can be well described as a random-walk in spin frequency, in which the variations are due to unresolved, random episodes of angular momentum transfer to the neutron star (Deeter et al. 1989 and references therein; Raubenheimer & Ögelman 1990).

From the Doppler shifts of the pulse arrival times, Rappaport et al. (1976) found that the orbit is moderately eccentric ($e \simeq 0.1$). This result has been confirmed by all subsequent studies of Vela X-1 (e.g., Rappaport et al. 1980; Van der Klis & Bonnet-Bidaud 1984; Boynton et al., 1986; Deeter et al. 1987a,b). In the present paper we have adopted the X-ray orbital parameters as derived by Deeter et al. (1987b; see Table 6).

A detailed optical radial-velocity study was made by Van Paradijs et al. (1977b), who derived an orbital eccentricity and periastron angle consistent with the corresponding values of the X-ray orbit. From the optical and X-ray mass functions, they inferred $M_X \sin^3 i = 1.67 \pm 0.12 \, M_\odot$ and $M_{\rm opt} \sin^3 i = 20.5 \pm 0.9 \, M_\odot$. Rappaport & Joss (1983; see also Joss & Rappaport 1984) reanalysed the orbital parameters using Monte-Carlo simulations to estimate the propagation of the observational errors on the resulting masses. They derived $M_X = 1.85^{+0.35}_{-0.30} \, M_\odot$ (95% confidence level error range). Nagase (1989) derived $M_X = 1.77^{+0.27}_{-0.21} \, M_\odot$ (90% confidence) applying the same technique on more recent X-ray orbital parameters and using the rotational velocity of HD 77581 derived by Sadakane et al. (1985).

The value of the mass of the neutron star mentioned above is of special interest, since it is the highest among all neutron-star mass determinations (Thorsett et al. 1993; Van Kerkwijk et al. 1995). If the value is indeed as high as the central value of the range indicates, it would strongly constrain the equation of state applicable for neutron-star matter, since for "softer" equations of state such a massive object collapses to a black hole (e.g., Arnett & Bowers 1977; Datta 1988; Stock 1989).

The uncertainty in the mass determination is mainly due to the relative inaccuracy of the optical data on HD 77581, especially the radial-velocity curve. However, since the time the last determination was made, the statistical and systematic accuracy of stellar spectroscopy has much improved, particularly by the introduction of CCD detectors. Because of the prospect offered by this improvement, we decided to obtain new spectra of the optical counterpart of Vela X-1, in an attempt to obtain a smaller error on the mass for this possibly quite massive neutron star. We find that this is hampered by large deviations from a purely orbital radial-velocity curve. In view of these deviations, the high accuracy of the velocity determinations became of rather limited use, and therefore we have also determined velocities for an older set of digitized photographic spectra and for the available IUE spectra. In this paper, we present the results of the analysis of all these spectra.

The observations and the data reduction are described in Sect. 2. In Sect. 3, we describe the cross-correlation technique used for the determination of the velocities. In Sect. 4, we present a Monte-Carlo technique with which we derive limits on the radial-velocity amplitude in the presence of excursions. Furthermore, we discuss possible causes for the large velocity excursions, and possible systematic effects that may influence the radial-velocity orbit. We use the limits on the radial-velocity amplitude in Sect. 5 to determine the masses of the two components in the system, and draw some conclusions in Sect. 6.

## 2. Observations and reduction

The present study is based on 40 echelle CCD spectra obtained in 1989, 13 nightly averages of digitized photographic spectra taken in 1975 and 1976, and 26 spectra obtained with the International Ultraviolet Explorer (IUE) in observing runs from 1978 to 1992. The observations are summarised in Tables 1–3. Below we discuss the reduction of the data in some detail.

### 2.1. CCD echelle spectra

We obtained 34 spectra of $\sim$ 45 minutes exposure each, covering the range 4190–4520 Å with a resolving power of $\sim$ 25 000, during five nights in February 1989. One more spectrum in the same wavelength region was kindly taken for us by Drs. W. Weiss and H. Schneider the night before our run, and five more, covering 4390–4715 Å, were obtained the five subsequent nights. All spectra were taken with the ECHELEC spectrograph, a Littrow echelle spectrograph with a grism as cross-disperser, mounted in a "white pupil" configuration at the coudé focus of the 1.52 m telescope at the ESO. The detector was a thinned, back-illuminated RCA CCD with 1024×640 pixels.

The spectra have been reduced using MIDAS supplemented with additional routines running in the MIDAS environment. The reduction procedure entailed the following steps: (i) subtraction of the electronic bias and the dark current ($\sim$ 30 electrons per pixel per hour, with a non-uniform distribution over the chip); (ii) finding defective columns and pixels and setting them to a value indicating 'undefined' for the subsequent steps; (iii) correction of 32 rows for offsets, which are constant at higher exposure levels, but approach zero non-linearly at low exposure levels (the dependence of the offsets on exposure level is determined using flat-field frames of different exposure times); (iv) subtraction of the diffuse background light, caused by scattering in the spectrograph (it is determined by interpolation between the inter-order values; see Verschueren & Hensberge 1990); and (v) global and local flat-field correction, detection and deletion of cosmic-ray events and order extraction.

For the last step an optimal extraction algorithm was developed, based on the method of Horne (1986), but generalised to allow for spectra not (closely) aligned with one direction on the chip (for details, see Van Kerkwijk 1993). Flat-field correction is part of the extraction process, since the algorithm needs the observed count rate in order to make an estimate of the error in each pixel. In the extraction process, cosmic-ray events are detected by the change they cause in the spatial distribution of star light. As a final check, we inspected the extracted orders by eye, substituting events that escaped detection with a linear interpolation between the adjoining pixels.

Despite the flat-field correction, the extracted orders showed a residual effect of the blaze of the grating. This is generally where $x$ and $y$ are the pixel number in the dispersion direction and the order number, respectively. The coefficients $a, \ldots, e$ are determined by the condition that the ratio between orders in the overlapping parts should be as close to unity as possible (by minimizing $\sum (Q-1)^2$, where $Q$ is the ratio between orders).

**Table 1.** CCD observations of HD 77581

| Sequence number[a] | JD$_{\text{mid. exp.}}$ $-2440000$ | Exp. time (min.) | Orbital phase[b] | Velocity[c,d] (km s$^{-1}$) |
|---|---|---|---|---|
| 1  | 7574.624 | 15 | 0.629 | $-3.1 \pm 1.4$ |
| 2  | 7575.527 | 15 | 0.730 | $-3.5 \pm 0.9$ |
| 3  | .551 | 45 | 0.732 | $-1.9 \pm 0.4$ |
| 4  | .591 | 45 | 0.737 | $0.0 \pm 0.5$ |
| 5  | .631 | 45 | 0.741 | $0.1 \pm 0.5$ |
| 6  | .676 | 45 | 0.746 | $1.8 \pm 0.5$ |
| 7  | .716 | 45 | 0.751 | $3.5 \pm 0.5$ |
| 8  | .800 | 45 | 0.760 | $7.5 \pm 0.6$ |
| 9  | .832 | 45 | 0.764 | $9.0 \pm 0.7$ |
| 10 | 7576.536 | 45 | 0.842 | $0.3 \pm 0.5$ |
| 11 | .584 | 45 | 0.847 | $-0.3 \pm 0.5$ |
| 12 | .636 | 45 | 0.853 | $0.2 \pm 0.5$ |
| 13 | .701 | 45 | 0.860 | $0.4 \pm 0.8$ |
| 14 | .736 | 45 | 0.864 | $0.2 \pm 0.7$ |
| 15 | .835 | 45 | 0.875 | $-0.4 \pm 0.9$ |
| 16 | .879 | 45 | 0.880 | $-1.9 \pm 1.1$ |
| 17 | 7577.530 | 45 | 0.953 | $0.7 \pm 0.5$ |
| 18 | .577 | 45 | 0.958 | $-0.1 \pm 0.5$ |
| 19 | .637 | 45 | 0.965 | $-0.2 \pm 0.6$ |
| 20 | .694 | 45 | 0.971 | $1.6 \pm 0.6$ |
| 21 | .727 | 45 | 0.975 | $2.5 \pm 0.6$ |
| 22 | .795 | 45 | 0.983 | $3.8 \pm 0.7$ |
| 23 | .872 | 23 | 0.991 | $6.3 \pm 1.6$ |
| 24 | 7578.541 | 45 | 0.066 | $25.2 \pm 0.5$ |
| 25 | .577 | 45 | 0.070 | $27.1 \pm 0.4$ |
| 26 | .627 | 45 | 0.075 | $26.6 \pm 0.5$ |
| 27 | .681 | 45 | 0.081 | $29.1 \pm 0.5$ |
| 28 | .721 | 45 | 0.086 | $28.4 \pm 0.5$ |
| 29 | .793 | 45 | 0.094 | $28.5 \pm 0.6$ |
| 30 | .847 | 45 | 0.100 | $29.0 \pm 0.7$ |
| 31 | 7579.525 | 45 | 0.175 | $45.2 \pm 0.7$ |
| 32 | .563 | 45 | 0.180 | $46.5 \pm 0.7$ |
| 33 | .617 | 45 | 0.186 | $47.4 \pm 0.7$ |
| 34 | .728 | 45 | 0.198 | $46.3 \pm 0.9$ |
| 35 | .840 | 60 | 0.211 | $44.1 \pm 0.9$ |
| 36 | 7580.529 | 30 | 0.287 | $1.0 \pm 0.9$ |
| 37 | 7581.542 | 30 | 0.401 | $-1.2 \pm 0.6$ |
| 38 | 7582.538 | 45 | 0.512 | $-32.6 \pm 0.8$ |
| 39 | 7583.591 | 30 | 0.629 | $-28.7 \pm 0.7$ |
| 40 | 7584.596 | 30 | 0.741 | $-30.8 \pm 0.8$ |

[a] Spectra 1–35 cover 4190–4520 Å, 36–40 cover 4390–4715 Å
[b] Using the ephemeris of Deeter et al. 1987b (see Table 6)
[c] Velocities are relative to spectrum 5 for spectra 1–35, and to spectrum 36 for spectra 36–40; all spectra are corrected for the shift observed in the interstellar lines using the fit with hour angle shown in Fig. 5 (due to this correction, the velocities of spectra 5 and 36 are not zero)
[d] Quoted are $1\sigma$ errors

**Table 2.** Photographic observations of HD 77581

| Plate number(s) | JD$_{\text{mid. exp.}}$ $-2440000$ | Total exp. t. (min.) | Orbital phase[a] | Velocity[b,c] (km s$^{-1}$) |
|---|---|---|---|---|
| G7115    | 2720.874 | 102 | 0.183 | $37.0 \pm 1.9$ |
| G7136–38 | 2721.842 | 93  | 0.290 | $37.9 \pm 1.0$ |
| G7158–64 | 2722.825 | 139 | 0.400 | $27.2 \pm 0.7$ |
| G7180–87 | 2723.832 | 103 | 0.512 | $7.0 \pm 0.7$ |
| G7201–08 | 2724.835 | 112 | 0.624 | $0.0 \pm 0.6$ |
| G7579–80 | 2908.557 | 38  | 0.119 | $32.4 \pm 1.5$ |
| G7591–94 | 2909.516 | 79  | 0.226 | $39.4 \pm 0.9$ |
| G7604–06 | 2910.518 | 70  | 0.338 | $31.1 \pm 1.3$ |
| G7617–19 | 2911.505 | 52  | 0.448 | $9.2 \pm 1.3$ |
| G7631–32 | 2912.626 | 109 | 0.573 | $-0.8 \pm 1.4$ |
| G7639–45 | 2913.554 | 202 | 0.676 | $3.1 \pm 0.8$ |
| G7654–57 | 2914.513 | 89  | 0.783 | $-1.6 \pm 0.6$ |
| G7668–77 | 2916.574 | 166 | 0.013 | $15.5 \pm 0.6$ |

[a] Using the ephemeris of Deeter et al. 1987b (see Table 6)
[b] Relative to G7201; corrected for shift of interstellar lines
[c] Quoted are $1\sigma$ errors

For wavelength calibration, Th/Ar frames were taken four or five times each night. These were reduced in the same way as the stellar spectra, except that the orders were extracted by directly adding a number of pixels around the centre of the order (after flat-field division), and that a possible residual blaze was not corrected for (because the slit is uniformly illuminated and because the light path is similar to that of the flat field). The positions of the lines were determined by taking the position of the maximum of the interpolating parabola of the three highest points of each line. Next, the lines were identified and a 2D-regression was made of wavelength as a function of pixel number and inverse order number. Typical r.m.s. residuals were 0.1 pixel ($\sim 5$ mÅ, corresponding to a velocity of $\sim 0.3$ km s$^{-1}$). During the individual nights the wavelength calibration remained the same to within 5 mÅ, while from night to night there were changes from 5 to 20 mÅ.

For each object frame, we used the wavelength calibration determined from the nearest (in time) Th/Ar frame or the average of the two nearest frames, to merge the orders and rebin them on a $\log \lambda$ scale (so that a velocity difference corresponds to a constant offset; see Sect. 3), with a bin size $\Delta\lambda/\lambda$ of $5\,10^{-6}$ ($\sim 20$ mÅ). Since the resolution elements of the instrument were much larger than the pixel size, the spectra were convolved with a Gaussian in order to remove high-frequency pixel-to-pixel noise. The width of the Gaussian was chosen such that the measured width of the interstellar lines ($\sim 0.2$ Å) did not increase by more than 10%.

The spectra were normalised through division by a second-degree polynomial fitted through selected line-free parts of the spectrum. In Fig. 1 a representative normalised spectrum is attributed to a slight difference in optical path between light from the star and the flat-field lamp. Following Gehren (1990), we corrected for this residual by dividing the extracted orders by a polynomial of the form $1 + ax + bx^2 + cxy + dx^2y + ex^2y^2$, shown for both wavelength intervals for which we have spectra, as well as the normalised averages of the spectra in the two sets. The wavelength regions used for the normalisation are given by Van Kerkwijk (1993) and indicated in the figure.

**Table 3.** IUE observations of HD 77581

| SWP number | $JD_{mid.\ exp.}$ $-2440000$ | Exp. time (min.) | Orbital phase[a] | Velocity[b,c] (km s$^{-1}$) |
|---|---|---|---|---|
| 1442 | 3628.707 | 180 | 0.453 | $0.0 \pm 2.2$ |
| 1488[d] | 3634.118 | 150 | 0.057 | $7.4 \pm 2.6$ |
| 2087[d] | 3712.994 | 125 | 0.856 | $-24.2 \pm 3.1$ |
| 3510[d] | 3845.206 | 90 | 0.604 | $-18.9 \pm 2.9$ |
| 3519[d] | 3846.188 | 140 | 0.714 | $-13.8 \pm 1.9$ |
| 3550[d] | 3850.013 | 140 | 0.140 | $11.4 \pm 2.8$ |
| 3649 | 3862.531 | 130 | 0.537 | $-14.4 \pm 2.7$ |
| 4718 | 3954.267 | 150 | 0.770 | $-16.4 \pm 2.3$ |
| 18823 | 5323.019 | 150 | 0.457 | $12.1 \pm 2.1$ |
| 18958 | 5341.590 | 180 | 0.529 | $1.5 \pm 2.1$ |
| 18970 | 5343.510 | 180 | 0.743 | $-6.4 \pm 2.3$ |
| 18983 | 5345.595 | 180 | 0.976 | $-2.3 \pm 2.0$ |
| 19012 | 5351.274 | 180 | 0.609 | $-19.7 \pm 2.2$ |
| 19061 | 5357.302 | 180 | 0.282 | $34.1 \pm 1.8$ |
| 22278 | 5746.814 | 150 | 0.733 | $-14.0 \pm 2.6$ |
| 22287 | 5747.819 | 150 | 0.845 | $-8.2 \pm 1.9$ |
| 22297 | 5748.989 | 125 | 0.975 | $-7.8 \pm 2.1$ |
| 22301 | 5749.818 | 138 | 0.068 | $6.2 \pm 2.8$ |
| 22309 | 5751.807 | 150 | 0.290 | $21.5 \pm 1.6$ |
| 22324 | 5752.856 | 150 | 0.407 | $5.1 \pm 2.2$ |
| 32961 | 7214.285 | 141 | 0.432 | $1.6 \pm 2.8$ |
| 32967 | 7215.282 | 141 | 0.543 | $-14.2 \pm 2.4$ |
| 33085 | 7233.269 | 130 | 0.550 | $-15.5 \pm 3.1$ |
| 46144 | 8933.056 | 180 | 0.165 | $20.3 \pm 1.9$ |
| 46151 | 8934.071 | 160 | 0.278 | $15.8 \pm 2.0$ |
| 46167 | 8935.216 | 165 | 0.406 | $10.5 \pm 2.1$ |

[a] Using the ephemeris of Deeter et al. 1987b (see Table 6)
[b] Relative to SWP 1442; corrected for shift of interstellar lines
[c] Quoted are $1\sigma$ errors
[d] Reprocessed with IUESIPS#2

### 2.2. Photographic coudé spectra

A subset of the photographic spectra used by Van Paradijs et al. (1977b) – the higher-quality baked-IIaO plates taken at 12 Å mm$^{-1}$ (resolving power $\sim 10\,000$) in observing runs in 1975 and 1976 – was digitized using the Faul-Coradi microdensitometer of the Astronomical Institute in Utrecht. For the reduction, the standard calibration procedures for photographically recorded spectra were followed (e.g., Underhill 1966; for details, see Zuiderwijk 1979). Wavelength calibration was performed using Iron-arc comparison spectra taken both before and after the stellar exposure. For the response curve, a simple analytical expression was used, whose parameters were determined from a fit to spectra obtained with a separate calibration spectrograph, using an incandescent light source and a rotating-step sector with 13 transmission steps. During the data processing, only the averages of spectra taken within one night were stored on magnetic tape, and these have been used here.

For the work described here, the spectra were rebinned to a logarithmic wavelength scale, with a resolution of $5\,10^{-6}$, the same as used for the echelle spectra (for convenience, but substantially oversampling the plate resolution). Furthermore, like the CCD spectra, the spectra were convolved with a Gaussian in order to remove high-frequency noise, with the width of the Gaussian chosen such that the measured width of the interstellar lines ($\sim 0.4$ Å) did not increase by more than 10%. The spectra were normalised by dividing by a seventh-degree polynomial fitted through selected parts of the continuum spectrum. The normalised spectrum from one of the higher-quality nightly averages and the normalised average of all spectra is shown in Fig. 1. The parts used for the normalisation are given by Van Kerkwijk (1993) and indicated in the figure.

### 2.3. IUE spectra

The IUE spectra were obtained with the SWP camera, and cover the wavelength range 1150–1950 Å. Some of the spectra have been published by Dupree et al. (1980) and Sadakane et al. (1985). These and additional spectra have been (re)analysed by Kaper et al. (1993). We refer to the latter authors for details about the reduction of the spectra, and for a figure showing a sample spectrum (the average obtained during X-ray eclipse).

For the work described here, the spectra were binned on a $\log \lambda$ scale with a resolution $\Delta\lambda/\lambda$ of $2\,10^{-5}$ ($\sim 35$ mÅ). The small data gaps in the spectra resulting from the reseaux pattern on the camera were set to a value "undefined". The spectra were normalised by first scaling them to the same flux level, and then dividing by a best-guess continuum obtained by interactively selecting points of the average spectrum, and interpolating these with a quadratic spline.

## 3. Velocity determination

We used cross-correlation of our spectra for the radial-velocity determination. The advantage of this technique is that one needs not worry about the intrinsic shape or symmetry of the lines, and whether they are blended or not. The only assumption made is that the template that is used to correlate the spectra with is a good model for those spectra, differing only in velocity. If that condition is fulfilled, the errors on the velocities can be derived in a straightforward way (similar to that used for a $\chi^2$-fit with one free parameter; see Van Kerkwijk 1993; also, e.g., Tonry & Davis 1979).

For cross-correlation purposes, the spectra should be normalised, and sampled on a logarithmic wavelength scale, so that a Doppler shift is a linear displacement through the whole spectral range. The velocity difference between two spectra is then found by fitting an analytic function to the peak of the discrete cross-correlation function (the correlation coefficient between the two spectra as function of velocity shift), and determining the position of the maximum of that analytic function. For our spectra, we found that the best results were obtained when we

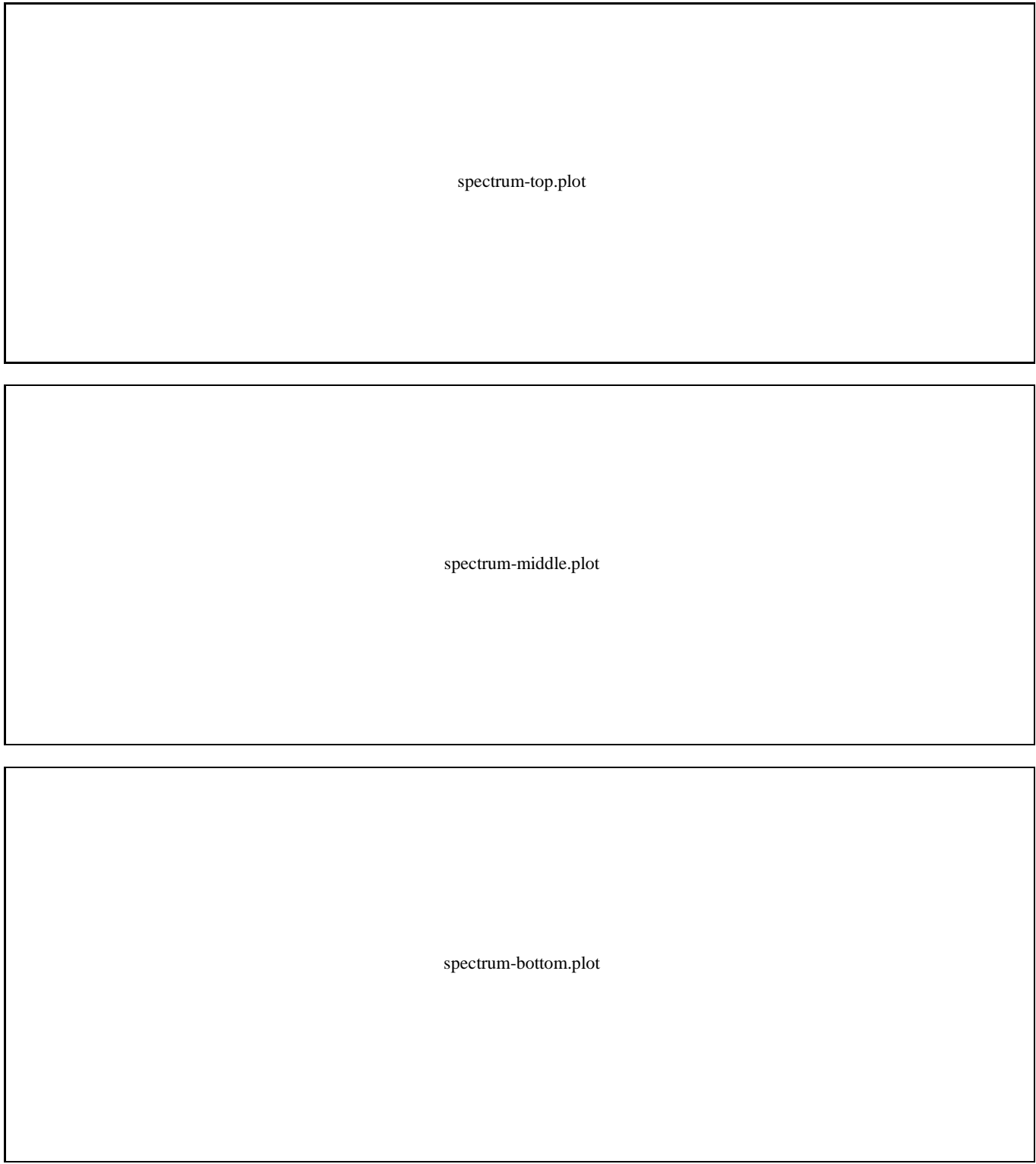

**Fig. 1.** The normalised optical spectrum of HD 77581 (B0.5Ib). Drawn are the average of the photographic spectra (a), photographic spectrum G7654 (b), the average of CCD spectra 1–35 (c), CCD spectrum 5 (d), the average of CCD spectra 36–40 (e) and CCD spectrum 37 (f). For the latter two, the wavelength region 4380–4475 Å is not shown. The vertical tickmarks in the figure are placed at 10% intervals. The wavelength regions used for the normalisation are shown at the top of each panel. For the line identifications, we used the identifications for $\epsilon$ Ori (B0Ia) by Lamers (1972), and those for HR 6165 (B0V) and HR 2928 (B2III) by Kilian et al. (1991). In case of blends, only the strongest component is indicated. The emission line at 4485 Å and the possible emission line at 4504 Å are also observed and possibly observed, respectively, in $\epsilon$ Ori (Lamers 1972; for a discussion on their origin, see Morrell et al. 1994). The interstellar bands are from Herbig (1975)

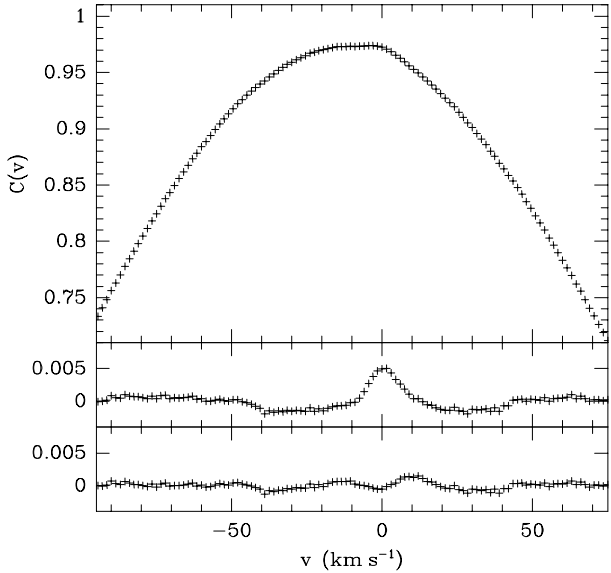 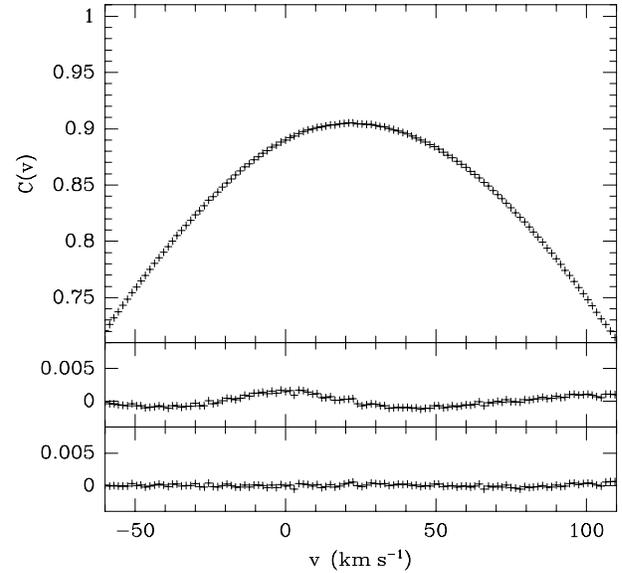

**Fig. 2.** The cross-correlation of a spectrum with an average as a template. In the upper panel the cross-correlation peak is shown of CCD spectrum 40 with the average of CCD spectra 36–40. Notice the excess peak at zero velocity due to the correlation of the noise in the spectrum with its diluted self in the average. In the middle panel the residuals with respect to a fit made with the sum of a Gaussian and a first-degree polynomial are shown. In the lower panel the same is shown for a fit with an extra Gaussian with its velocity fixed at zero and its width at $5\,{\rm km\,s^{-1}}$. The velocity difference between the two fits is $0.2\,{\rm km\,s^{-1}}$

**Fig. 3.** Same as Fig. 2, but for the cross-correlation of photographic spectrum G7115 with the average of the 13 photographic spectra. Notice that due to the lower resolution the excess peak at zero velocity is broader than in Fig. 2. The velocity difference between the two fits is $0.6\,{\rm km\,s^{-1}}$

fitted the sum of a Gaussian and a first-degree polynomial to the top 30% of the correlation peak.

Because of the condition of similarity of template and spectrum, we at first decided to use the average of each set of spectra taken in the same way as the template for that set. However, this choice leads to a complication, viz. that there will be a small excess peak at zero velocity superposed on *all* correlation peaks (see Fig. 2, 3), because the noise in each spectrum is also present in a diluted form in the average, and thus correlates with itself (see Van Kerkwijk 1993 for an analysis of the expected behaviour of this autocorrelation of the noise). Clearly, if this peak lies on the part of the correlation peak that is used to determine the velocity, then the derived velocities will be systematically drawn to zero. For our photographic spectra, for instance, we found that the velocities were biased towards zero by up to $0.6\,{\rm km\,s^{-1}}$ ($\sim 3\%$). This effect can be corrected for by either excluding the points around zero velocity from the fit of the cross-correlation peaks, or by including an extra Gaussian in the fit with its centre fixed to zero and its width determined by the size of independent elements, i.e., the (filtered) pixel size. The latter is easily determined from autocorrelations of the spectra (from the autocorrelations, we also found that for our spectra a Gaussian is indeed a good model for the excess peak). Alternatively, instead of using the average as the template, one can cross-correlate all the spectra with each other, and determine the individual velocities from the velocity differences (with one velocity fixed to zero).

For the CCD spectra and the photographic spectra we tried both cross-correlating all spectra with each other and including an extra Gaussian at zero velocity. We found that the derived velocities were the same to within $0.8\,{\rm km\,s^{-1}}$ for the CCD spectra, and to within $0.2\,{\rm km\,s^{-1}}$ for the photographic spectra (corresponding to a linear dependence deviating by less than 2 and 0.5% from unity, respectively).

For the IUE spectra, the situation is somewhat more complicated, because independently of the chosen wavelength range there always remains an excess peak superposed on many of the cross-correlation peaks (see Fig. 4). This effect, previously reported by Evans (1988), most likely results from the so-called 'fixed-pattern noise' in the spectra, which is due to inadequately corrected pixel-to-pixel variations in sensitivity. We decided to correct for this by including an extra Gaussian in the fit which has only its width fixed. Fitting the cross-correlation peaks of spectra with the average becomes rather difficult, since now two excess peaks should be accounted for, one for the systematic pixel-to-pixel variations, and one for the autocorrelation of the noise. Given the consistent results obtained from the optical spectra, we decided not to attempt this, but rather to use the velocities derived from the cross-correlations of all spectra with each other.

For each of the data sets, we determined radial velocities for two sets of wavelength regions (listed in Table 4), one with small regions containing the interstellar lines and one with the remainder of the spectral region (excluding H$\beta$ in the optical, for which Van Paradijs et al. (1977b) found that it showed strong deviations from the other lines, and the wind resonance lines in the UV). We also determined stellar velocities for each of the individual wavelength regions listed in Table 4, as well as for several smaller regions centred on individual lines. We found that all these were entirely consistent.

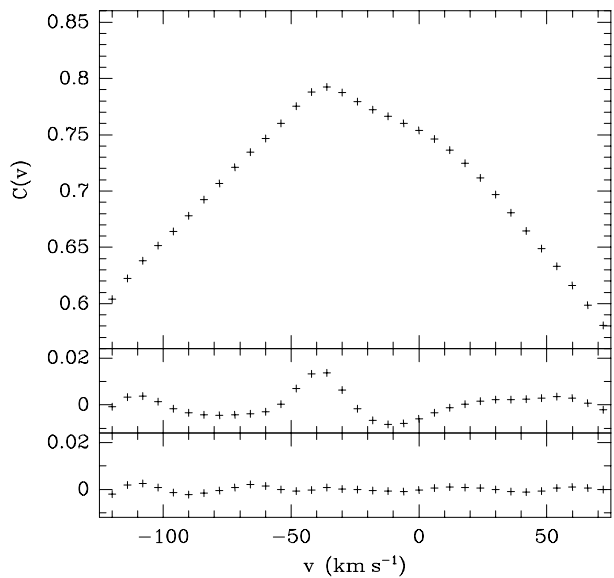

**Fig. 4.** The cross-correlation of two IUE spectra. In the upper panel the cross-correlation peak is shown of spectrum SWP 3550 with SWP 32967. Notice the large excess peak at $-40\,\mathrm{km\,s^{-1}}$ that is probably due to the correlation of the fixed-pattern noise in the two spectra. In the middle panel the residuals with respect to a fit made with a single Gaussian are shown. In the lower panel the same is shown for a fit with an extra Gaussian with its width fixed at $25\,\mathrm{km\,s^{-1}}$. The velocity difference between the two fits is $1.5\,\mathrm{km\,s^{-1}}$

The velocity differences for the interstellar lines are not all zero. For CCD spectra 1–35, the deviations vary linearly with hour angle (see Fig. 5; $\chi^2 = 37$ for 33 degrees of freedom). Therefore, we corrected the stellar velocities for the CCD spectra using this relation. We also applied the correction to spectra 36–40, in which no interstellar lines are present, but which were taken with the same instrumental setup except for the change in central wavelength (notice that the range in hour angle spanned by spectra 36–40 is small; hence, the results hardly depend on whether we apply the correction or not). For the photographic and IUE spectra, each stellar velocity was simply corrected with the interstellar velocity determined for that spectrum. All corrected stellar velocities are listed in Tables 1–3.

## 4. The radial-velocity orbit

The stellar velocities determined from all spectra are shown as a function of orbital phase in Fig. 6. From this figure, it is obvious that the velocities deviate substantially from the smooth radial-velocity curve expected for pure Keplerian motion. Within one night, the deviations are autocorrelated (see the CCD spectra), while from one night to the other they are not. Similar excursions within one night were already noticed by Van Paradijs et al. (1977b), and, with hindsight, seem to be present as well in the data of Hiltner et al. (1972) and Zuiderwijk et al. (1974).

**Table 4.** The wavelength regions used for the velocity determination

| Optical[a] | | | Ultraviolet | | |
|---|---|---|---|---|---|
| $\lambda_{\mathrm{start}}$ (Å) | $\lambda_{\mathrm{end}}$ (Å) | Main line(s) | $\lambda_{\mathrm{start}}$ (Å) | $\lambda_{\mathrm{end}}$ (Å) | Main line(s) |
| *Interstellar* | | | *Interstellar* | | |
| 3932.0 | 3936.0 | Ca II K | 1560.2 | 1560.9 | C I |
| 4231.0 | 4235.0 | CH$^+$ | 1608.3 | 1608.7 | Fe II |
| 4298.5 | 4302.5 | CH | 1656.2 | 1657.7 | C I |
| *Stellar* | | | 1807.7 | 1808.3 | Si II |
| 3700.0 | 3932.0 | H$_{8-16}$, He I | *Stellar* | | |
| 3974.0 | 4231.0 | H$_\delta$, He I | 1563.0 | 1607.0 | Fe IV, C III |
| 4235.0 | 4298.0 | S III | 1611.0 | 1655.0 | Fe IV, He II |
| 4303.0 | 4400.0 | H$_\gamma$, He I | 1672.0 | 1707.0 | Fe IV |
| 4450.0 | 4476.0 | He I | 1729.0 | 1741.0 | Fe IV, N III |
| 4500.0 | 4582.0 | Si III | 1745.0 | 1806.0 | N III, O III, |
| 4586.0 | 4690.0 | O II, C III | | | Fe IV, Cr IV |
| 4696.0 | 4720.0 | He I | 1810.0 | 1842.0 | Fe IV, Cr IV |

[a] For the photographic spectra all regions are used; for CCD spectra 1–35 the regions from 4231 to 4476 Å, and for CCD spectra 36–40 the regions from 4450 to 4690 Å

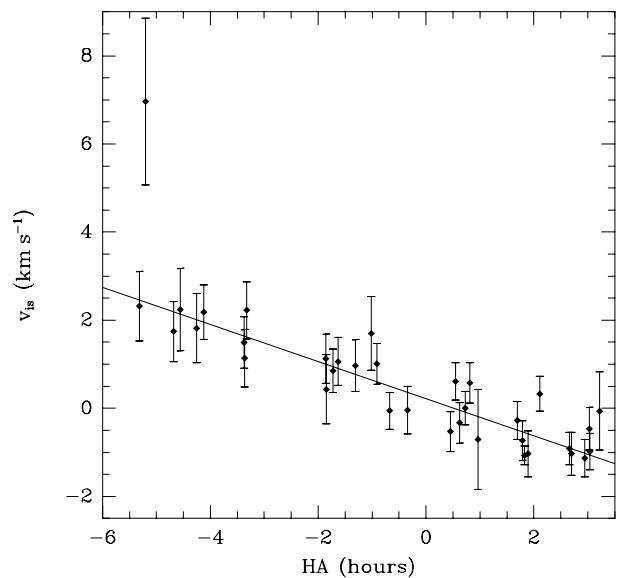

**Fig. 5.** The velocities derived from the interstellar lines in CCD spectra 1–35 as function of hour angle. Overdrawn is a linear fit, which has $\chi^2 = 37$ for 33 degrees of freedom

### 4.1. The uncertainty introduced by the velocity excursions

Due to the presence of autocorrelated deviations, not all velocity determinations are independent. Hence, a direct $\chi^2$-fit of a radial-velocity orbit to the data to determine the parameters and the corresponding uncertainties, is not meaningful. For this reason, Van Paradijs et al. (1977b) made a $\chi^2$-fit to the averages of the velocities obtained during one night. A problem with this approach is that there is no obvious way to obtain a reasonable estimate of the errors on the nightly averages. As can be seen in Fig. 6, observations during one night do not fully cover an excursion, and hence the standard deviation around the mean in each night is not necessarily a good estimate of the error. Also, there is, a priori, no reason to assume a normal distribution of the deviations with respect to the radial-velocity orbit.

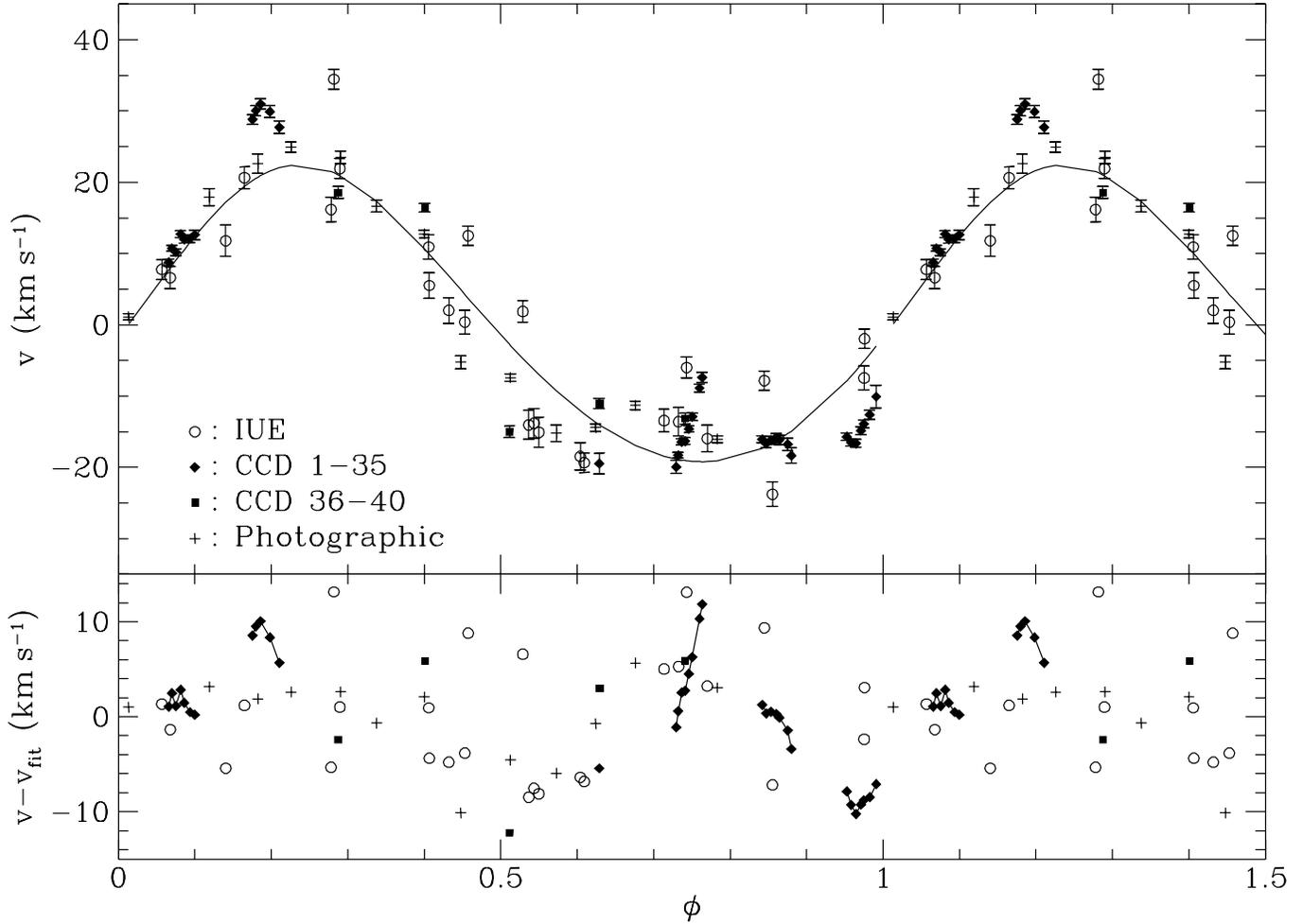

**Fig. 6.** The stellar velocities derived for all spectra as a function of orbital phase (using the ephemeris of Deeter et al. 1987b (see Table 6); note that $\phi = 0$ corresponds to time of mean longitude $\pi/2$; periastron is at $\phi = 0.17$ (see Van der Klis & Bonnet-Bidaud 1984)). The error bars indicate the $1\sigma$ uncertainties. The velocities in each data set have been shifted to zero system velocity based on the best-fitting (minimal root-mean-square deviation) radial-velocity orbit. The latter is overdrawn. In the lower panel the residuals are shown. For clarity, the error bars have been omitted. Points connected by lines were taken within one night

To get an idea of the effect of the excursions on the accuracy of the determination of the orbital elements, we performed Monte-Carlo simulations, in which artificial sets of velocity data were generated using the (smooth) best-fit radial-velocity orbit plus a model for the velocity excursions. A velocity was generated for each epoch of the real data, and a 'measurement' error added, randomly drawn from a normal distribution with a standard deviation corresponding to the real measurement error. This approach allows an exploration of the effects of different models for the excursions and the consequences of different fitting methods, and thus to search for the model that maximizes the range in parameters found (while statistically producing the same kind of velocity sets as observed) and the fitting method that minimizes it (depending as little as possible on the model for the excursions). The only implicit assumption is that the excursions do not depend on orbital phase, i.e., that no systematic orbital-phase dependent effects are present (we will return to this in Sect. 4.3).

The velocity excursions within one night can be described well with a linear or quadratic function of time (Fig. 6). In order to reproduce this behaviour, we modelled the excursions with sinusoids whose frequency, amplitude and initial phase are drawn randomly from given distributions for every night on which observations were taken (remaining fixed throughout a night). As expected, the period of the sinusoids has to be longer than one day, so that in any night the fraction of the sinusoid sampled is small enough to obtain the observed approximately linear or quadratic form.

Three different fitting methods were tested: (i) minimizing the mean absolute deviation; (ii) minimizing the root mean square deviation; and (iii) minimizing $\chi^2$. For all three, the velocities are fitted with a radial-velocity curve whose orbital period $P_{\rm orb}$, time $T_{\pi/2}$ of mean longitude $\pi/2$, eccentricity $e$ and periastron angle $\varpi_{\rm opt}$ are fixed at the values obtained from the analysis of the X-ray data by Deeter et al. (1987b; see Table 6; $\varpi_{\rm opt} = \varpi_{\rm X}+180°$). The remaining free parameters are the radial-velocity amplitude $K_{\rm opt}$ and the system velocity $\gamma$ for each data set (since only relative velocities in each data set are available). Of the three fitting methods, only the third takes into account the measurement errors. Since these clearly do not correspond to good estimates of the real deviation from the radial-velocity curve, one expects a priori that this method will give rather inaccurate results. However, the range should include the results from the other two methods. Similarly, since observations taken in one night show autocorrelated deviations, it seems reasonable to give those less weight. Therefore, additional simulations were made in which both the real and the simulated data are averaged within nights before being fitted.

The amplitude, frequency and initial phase of the sinusoidal velocity excursions are drawn from a population distributed according to the form $C_0 + C_u u + C_n n$, where $u$ is a uniform distribution in the range $0\ldots 1$, $n$ a normal distribution with mean zero and standard deviation unity, and $C_0$, $C_u$ and $C_n$ are constants. Obviously, there is no reason to assume that the phase at the beginning of a night has any preferred value. Hence, $C_{\phi,0} = 0$, $C_{\phi,u} = 2\pi$, and $C_{\phi,n} = 0$ were used throughout.

Results of the simulations are listed in Table 5 for three different types of distribution of the amplitude: constant, uniform, and normal. For each distribution, the appropriate constant was chosen such that the average mean absolute deviation, root-mean-square or $\chi^2$ from the simulations was the same as the observed one. By way of verification, for all fitting methods the cumulative distribution of the deviations of the observed velocities from the fitted radial-velocity orbits was determined, and compared with the average cumulative distributions from the Monte-Carlo simulations. The observations could be reproduced with all three distributions of the amplitude.

The frequency distribution is reflected mostly in the rate of change of the velocity deviations. The latter can be estimated to first order by $\Delta(v - v_{\rm fit})/\Delta JD$, where $\Delta(v - v_{\rm fit})$ is the difference in deviation from the fitted radial velocity orbit for two subsequent observations, and $\Delta JD$ the corresponding time interval. A good match to the observed distribution of values for the rate of change of the velocity deviations can be obtained if the mean frequency is about $0.7\,\rm day^{-1}$, consistent with the fact that excursions do not seem to be covered well in one night (values of $\gtrsim 1\,\rm day^{-1}$ give significantly worse results). The results are rather insensitive to the exact form of the frequency distribution. Therefore, in Table 5 only the results obtained for a constant value of the frequency are listed (chosen such that the observed cumulative distribution of the rate of change was best reproduced).

From Table 5 it is clear that minimizing the root-mean-square deviation provides the best results, both in the sense that the range in radial-velocity amplitude is minimal, and in the sense that the result depends least on the chosen form of the distribution of the excursions. As expected, the range in radial-velocity amplitude found is somewhat smaller when velocities are averaged within nights. For this case, the radial velocity amplitude is given by $K_{\rm opt} = 20.8 \pm 1.4\,\rm km\,s^{-1}$ $(1\sigma)$.

**Table 5.** Results of the Monte-Carlo simulations

| Ampl. distr.[a] (km s$^{-1}$) | Freq. distr.[a] (day$^{-1}$) | Fit type[b] | All data Fit val.[b] | $K_{\rm opt}$[c] (km s$^{-1}$) | Nightly averages Fit val.[b] | $K_{\rm opt}$[c] (km s$^{-1}$) |
|---|---|---|---|---|---|---|
| *Real data* | | mad | 4.65 | 21.7 | 4.76 | 21.7 |
| 8.4 | 0.6 | | 4.70 | 16.8–26.8 | 4.87 | 17.6–25.9 |
| 15.5$u$ | 0.7 | | 4.67 | 17.8–25.8 | 4.79 | 18.6–24.9 |
| 9.5$n$ | 0.7 | | 4.65 | 18.1–25.4 | 4.76 | 18.9–24.5 |
| *Real data* | | rms | 7.28 | 21.5 | 6.47 | 20.8 |
| 9.0 | 0.6 | | 7.33 | 18.1–25.0 | 6.64 | 18.2–23.5 |
| 15.4$u$ | 0.7 | | 7.28 | 18.1–24.9 | 6.59 | 18.2–23.4 |
| 9.0$n$ | 0.7 | | 7.31 | 18.0–25.0 | 6.61 | 18.2–23.5 |
| *Real data* | | $\chi^2_{\rm red}$ | 52.6 | 21.4 | | |
| 9.0 | 0.6 | | 52.7 | 15.7–27.2 | | |
| 15.4$u$ | 0.7 | | 52.5 | 15.7–27.2 | | |
| 8.9$n$ | 0.7 | | 53.0 | 15.5–27.3 | | |

[a] $u$, $n$ indicate uniform distribution $0\ldots 1$ and normal distribution with mean zero and standard deviation unity, respectively.
[b] The quantity which is minimized: mean absolute deviation (mad), root mean square (rms), or reduced chi-square ($\chi^2_{\rm red}$). The value is given under column Fit val. The unit is km s$^{-1}$ for mad and rms, and dimensionless for $\chi^2_{\rm red}$
[c] Best fit value for the real data, 95% confidence region for the simulations (with the best-fit value as input radial-velocity amplitude).

### 4.2. Line profile variations

To improve the understanding of the cause of the deviations, we show the profiles of He I $\lambda 4471$, corrected for the orbital motion with the minimal root-mean-square solution, for all CCD spectra in Fig. 7 (left-hand panel). Clearly, there are substantial changes in the shape of the profile. The timescale of these variations is longer than one night, as is expected from the fact that the velocity excursions do not seem to be covered within one night (see Fig. 6), as well as from the results of the Monte-Carlo simulations. However, there is no clear correlation between the shape of the lines and the excursions. For instance, during the night that the largest excursion occurred, the line shape hardly changed (2nd to 9th curve in Fig. 7). The line shape is not a strict function of orbital phase either: at the two orbital phases for which we have more than one point, different line shapes are observed (compare the bottom curve with that second from the top, and the 2nd to 9th curves with that at the top).

To show the changes from night to night more clearly, the average He I $\lambda 4471$ profiles for the five nights for which we have more than one spectrum are shown in the right-hand panel of Fig. 7. Also shown are the average profiles of He I $\lambda 4388$, H$\gamma$ and S III $\lambda 4254$. From the figure, it is clear that the profile changes are very similar for all lines. This suggests that they are not due to phenomena related to the wind, such as density enhancements in an accretion or ionisation wake, since these are expected to show up preferably in the strongest lines. This is consistent with the fact that the velocities derived for individual lines are all consistent with those derived for the whole spectrum.

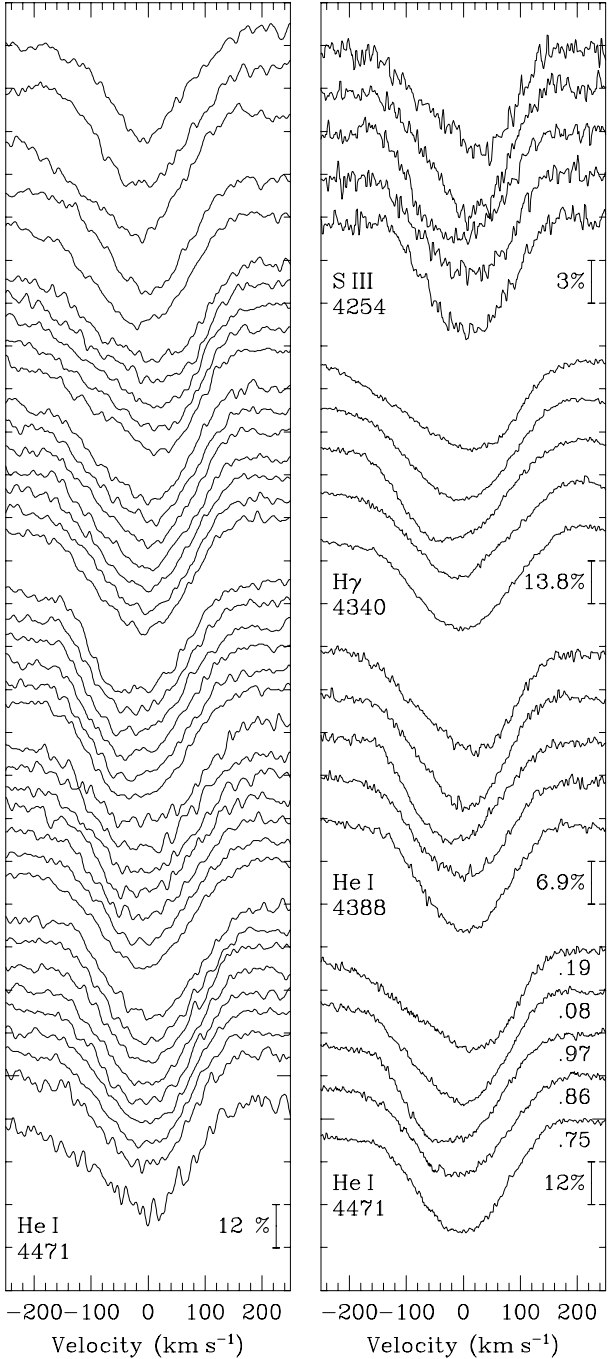

**Fig. 7.** Line profile variations. In the left-hand panel the line profile of He I λ4471 is shown for all CCD spectra. The spectra have been shifted vertically, time increasing upward (nights being separated by somewhat larger shifts). The profiles are shifted to the rest frame of the optical star using the minimal root-mean-square orbital solution (see Fig. 6). In the right-hand panel, the average profiles of He I λ4471, He I λ4388, Hγ and S III λ4254 are shown for the five nights for which we have more than one spectrum. The orbital phase is indicated for He I λ4471. To ease comparison, the profiles have been scaled to the same relative depth. Notice that He I λ4471 is blended on the short-wavelength side and Hγ on both sides (see Fig. 1). Also, S III λ4254 contains a contribution from O II lines (at approximately the same wavelength)

Instead, the similarity in the profiles seems to indicate that the variations in shape and the changes in velocity reflect large-scale motion of the surface. The underlying physical mechanism might be that proposed by Tjemkes et al. (1986) to explain the irregular variations seen in the optical light curve (which occur at very similar timescales), viz., that the star is pulsating with many modes (possibly excited by the varying tidal force exerted by the neutron star in its eccentric orbit), which for short periods of time interfere constructively, leading to quasiperiodic oscillations.

### 4.3. Possible systematic effects

Although the velocity differences with respect to the orbital fit are dominated by more or less random excursions (Fig. 6), it is worthwhile to investigate whether there are any lower-amplitude, but systematic effects hidden in the data. We will discuss both effects due to the tidal deformation of the star and effects occurring at particular orbital phases, e.g., periastron passage or inferior conjunction of the compact object.

#### 4.3.1. Systematic effects due to tidal deformation

The presence of ellipsoidal variations in the optical light curve shows that HD 77581 is deformed due to the tides induced by its companion. On the basis of numerical calculations of the line profiles of a deformed star, Van Paradijs et al. (1977a) found that the deformation is expected to induce phase-locked deviations of a few $\mathrm{km\,s^{-1}}$ in the observed radial velocities, with corresponding changes in the derived $e$ and $K_{\mathrm{opt}}$ of $\lesssim 0.06$ and a few $\mathrm{km\,s^{-1}}$, respectively (see also Van Paradijs et al. 1977b; Hutchings 1977; Wilson 1979). The amplitude of the velocity deviations is expected to depend on the line that is studied, but neither for the photographic spectra (Van Paradijs et al. 1977b), nor for the CCD spectra this is found. If we leave $e$ and $\varpi_{\mathrm{opt}}$ as free parameters in the orbital fit, we find a change in eccentricity that is of the expected order: $e = 0.21$, $\varpi_{\mathrm{opt}} = 359°$ and $K_{\mathrm{opt}} = 21.5\,\mathrm{km\,s^{-1}}$. However, it follows fits to the data sets generated in the Monte-Carlo simulations that the change is significant with 90% confidence only.

Although the evidence for tidal effects is not strong, we stress that that we cannot rule it out. In fact, it seems unlikely that the deformation does not influence the radial velocities, while it is reflected clearly in the ellipsoidal variations. Since the amplitude of the light curve can be modeled reasonably well (though not the exact shape; see Tjemkes et al. 1986), it seems likely that the expected deviation of $K_{\mathrm{opt}}$ gives a reasonable measure of the uncertainty introduced by the tidal effects.

As mentioned above, for Vela X-1 changes in the $K_{\mathrm{opt}}$ of a few $\mathrm{km\,s^{-1}}$ are expected (Van Paradijs et al. 1977a,b; Hutchings 1977; Wilson 1979). Other estimates for the tidal effect in massive X-ray binaries have been made for SMC X-1 (without X-ray heating, Wilson & Sofia 1976) and for 4U 1538-52 (Reynolds et al. 1992). Recently, model calculations have been used to correct radial-velocity curves of W Ursa Majoris binaries and Algols (Hill et al. 1989; Khalesseh & Hill 1992 and references therein; see also Van Hamme & Wilson 1985). Although these binaries contain components of later spectral type, the corrections that are found are similar to what is found for the massive X-ray binaries: about 5% (for both components of the binary). Therefore, a reasonable estimate of the '$1\sigma$' systematic uncertainty seems to be given by a relative error of 5%, smaller than the error introduced by the velocity excursion, but not negligible.

### 4.3.2. Systematic effects at particular orbital phases

In previous observations of HD 77581, systematic excursions at two particular phases have been reported. In the observations of Hiltner et al. (1972) and Zuiderwijk et al. (1974), large positive excursions around velocity minimum (phase 0.75) are seen, very similar to what is shown by our CCD spectra. A smaller excursion at the same phase seems to be present in the data of Van Paradijs et al. (1977b). From Fig. 6, also the IUE data seem to deviate most systematically from zero at velocity minimum. This suggests that there may be a recurring phenomenon.

Another phase interval at which systematic effects may be present, is that shortly following inferior conjunction of the X-ray pulsar ($\phi \simeq 0.5$), when outflowing density enhancements in the line of sight could influence the velocities. Such enhancements have been inferred from variations of H$\alpha$ (e.g., Zuiderwijk et al. 1974), and from various features in X-ray light curves and spectra (Watson & Griffiths 1977; Nagase et al. 1986; Haberl & White 1990). Also, for our photographic spectra, Van Paradijs et al. (1977b) found that near phase 0.5 the H$\beta$ velocities deviated systematically from those for the other lines. From the H$\beta$ line profiles, it is clear that this is due to an extra absorption component, which moves from a low velocity at phase 0.5 to a velocity of about $-300\,\mathrm{km\,s^{-1}}$ at phase 0.9, at which it disappears (Zuiderwijk 1979, curve b in Fig. 1). Kaper et al. (1994) find similar profile variations for He I $\lambda$4471, which they attribute to a photoionisation wake trailing the X-ray source. While such variations are not seen in our CCD spectra (Fig. 7), we cannot exclude their presence in the photographic or the IUE data.

In order to investigate the influence of possible orbital-phase related effects, we made fits to the radial velocities excluding certain orbital phase intervals, of size 0.1 and size 0.2. The results are shown in Fig. 8. From this figure, it is clear that around phase 0.75 a large deviation occurs, while any effect following inferior conjunction is relatively small.

In order to determine the significance of the deviation, we also made orbital fits with phase intervals excluded for the data sets generated in the Monte-Carlo simulations. From these calculations the ranges containing 95% and 99% of the derived values of $K_{\rm opt}$ were determined. The results (Fig. 8) indicate that the largest effects can be expected when excluding intervals close to the phases where the velocity reaches an extreme (the curve is not very smooth, especially in the upper panel, be-cause the number of data points actually excluded differs for the different excluded phase intervals; e.g., for the interval 0.3–0.4 only one point is excluded).

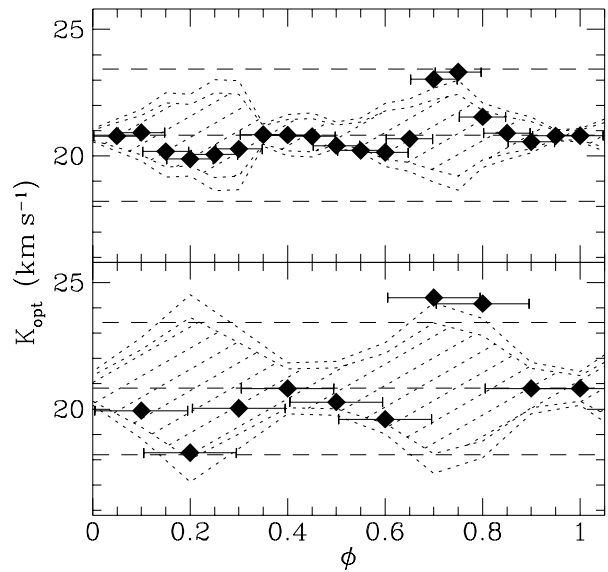

**Fig. 8.** Radial-velocity amplitude found when orbital phase intervals of size 0.1 (top) and 0.2 (bottom) are excluded. The points indicate the results of fits to the observed velocities and the short-dashed lines the envelopes which contain 95% (inner region; shaded) and 99% (outer) of the results for the Monte-Carlo simulations. The long-dashed lines indicate the central value and 95% confidence limits derived for all data

The simulations revealed that the observed deviation at $\phi = 0.75$ is the most significant, even though the expected deviation at that phase is rather large. To quantify the significance, one has to take into account that effectively 30 'trials' were made to find a deviating radial-velocity amplitude (where those 30 trials are not independent). We did this by determining the fraction of Monte-Carlo simulations that was more deviant than the observed data. As a measure of deviation a statistic $D^2 = \frac{1}{n}\sum_{i=0}^{n}(K_i^{\rm x} - K^{\rm all})^2/\sigma_i^2$ was used, where $K^{\rm all}$ and $K_i^{\rm x}$ are the radial-velocity amplitudes for all velocities and for velocities within bin $i$ excluded, respectively, and $\sigma_i$ is the root-mean-square deviation expected from the Monte-Carlo simulations. For the case where intervals of size 0.1 were excluded, $D^2_{\rm obs} = 1.90$, which is larger than the $D^2$ of 96.4% of the simulations. For the intervals of size 0.2, $D^2_{\rm obs} = 2.72$, larger than the $D^2$ for 98.5% of the simulations. Thus, it appears that the systematic deviation around $\phi = 0.75$ is significant at at least the 95% confidence level, suggesting that at this position the velocity excursions do correlate with orbital phase, perhaps because here some distortion effect repeats itself in every orbit.

It was mentioned above that if $e$ and $\varpi_{\rm opt}$ are left free in the orbital fit to all data, values not quite consistent with those derived from X-ray timing are found. The same is true for fits with phase intervals excluded, except for intervals around phase 0.75. With the interval 0.65–0.85 excluded, e.g., we find $e = 0.11$ and $\varpi_{\rm opt} = 340°$, consistent with the X-ray results.

*4.4. The final orbital parameters*

From the above, we conclude that the accuracy with which the orbital parameters can be determined is limited by: (i) velocity excursions that are autocorrelated within one night, but not from night to night; (ii) possible effects due to tidal deformation; and (iii) the possible presence of systematic positive deviations in velocity close to the time of velocity minimum. While from our Monte-Carlo simulations and from published model calculations we can estimate the uncertainties introduced by (i) and (ii), respectively, we do not know how to estimate the uncertainty related to (iii). From our simulations, the effect seems significant at the $\gtrsim 95\%$ confidence level. However, in the absence of a physical mechanism to explain why deviations occur preferentially around $\phi = 0.75$ (the closest 'relevant' orbital phase is apastron, at $\phi = 0.68$), we feel hesitant to adopt the results found when the velocities around $\phi = 0.75$ are excluded. The most conservative approach seems to be to adopt a range in radial-velocity amplitude of 18.0–28.2 km s$^{-1}$, defined by the 95% confidence lower limit for the fit to all data and the 95% confidence upper limit for the fit with phase 0.65–0.85 excluded (for both limits, the uncertainty due to tidal effects is combined quadratically with that due to the velocity excursions). For comparison, we will also list results for the two cases individually, i.e., $K_{\rm opt} = 20.8 \pm 1.7$ km s$^{-1}$ (1$\sigma$) for all data, and $K_{\rm opt} = 24.6 \pm 2.3$ for phase 0.65–0.85 excluded.

## 5. The masses of the two components

The limits on the radial-velocity amplitude $K_{\rm opt}$ derived in the previous section, in combination with the orbital period $P_{\rm orb}$, eccentricity $e$ and semi-major axis $a_{\rm X} \sin i$ of the X-ray orbit of Vela X-1 (see Table 6), can be used to derive limits on the masses of the two components (for formulae, see e.g. Rappaport & Joss 1983; Nagase 1989). For Vela X-1, we find, using the range in $K_{\rm opt}$ found above, $M_{\rm X} \sin^3 i = 1.42$–$2.38\,M_\odot$ and $M_{\rm opt} \sin^3 i = 21.8$–$23.3\,M_\odot$ (where $i$ is the inclination of the system).

If one had an estimate of the radius $R_{\rm opt}$ of the companion, the inclination $i$ could be estimated from the duration of the X-ray eclipse $\theta_{\rm ecl}$. For Vela X-1, the optical light curve shows strong ellipsoidal variations, which indicates that the optical companion is close to filling its Roche lobe (e.g., Tjemkes et al. 1986). For a circular orbit and a corotating optical companion, this would mean that the radius of the star could be estimated with the radius of a sphere with a volume close to the volume of the Roche lobe, and hence that its size relative to the orbital separation would be a function of mass-ratio $q$ and filling fraction $\beta$ only. For a non-corotating star and an eccentric orbit like for Vela X-1, generalisations of the Roche potential have been derived by Avni & Bahcall (1975) and Avni (1976) under the assumption that the star is in quasi-hydrostatic equilibrium at any given time with respect to the instantaneous potential. Using these generalisations, one can estimate the 'Roche' radius as a function of orbital phase for given mass-ratio $q$, orbital eccentricity $e$, periastron angle $\varpi$, and corotation factor $f_{\rm co}$ (which can be derived from the measured rotation velocity $v_{\rm rot} \sin i$ of the optical star), and hence the inclination for a given filling factor $\beta$ at periastron.

Rappaport & Joss (1983; also Joss & Rappaport 1984; Nagase 1989; Van Kerkwijk et al. 1995) have used this method to estimate the inclination, the semi-major axis, the radius and mass of the companion, and the neutron-star mass for Vela X-1 and for the five other X-ray binaries for which all necessary parameters are known. The corresponding uncertainties are found by means of a Monte-Carlo error propagation technique, in which a large number of trial evaluations are made for values of $P_{\rm orb}$, $e$, $\varpi$, $a_{\rm X} \sin i$, $\theta_{\rm ecl}$, $K_{\rm opt}$, $f_{\rm co}$ and $\beta$ drawn from random distributions which reflect the experimental and theoretical uncertainties. These distributions are either normal or uniform between approximate 99% confidence limits.

We have applied this method for Vela X-1, using the X-ray orbit of Deeter et al. (1987b) and three different distributions of the radial-velocity amplitude, viz., uniform between the 99% lower limit derived from the fit to all data and the 99% upper limit derived from the fit with phase 0.65–0.85 excluded, and the individual solutions for the two cases. For $\theta_{\rm ecl}$, we conservatively used a range 30°–36°, which encompasses the values quoted by Watson & Griffiths (1977; 33°.8 $\pm$ 1°.3, ARIEL V), Nagase et al. (1983; 32° $\pm$ 1°, Hakucho 9–22 keV) and Sato et al. (1986; 34°.4 $\pm$ 1°.1, Tenma 10–20 keV).

Based on a comparison of the photographic spectra with model profiles calculated for a deformed star, Zuiderwijk (1995) finds $v_{\rm rot} \sin i = 115 \pm 6$ km s$^{-1}$. From the CCD spectra, we estimate a value of $110 \pm 15$ km s$^{-1}$. Estimates based on photographic spectra were 130 km s$^{-1}$ (Mikkelsen & Wallerstein 1974) and 90 km s$^{-1}$ (Wickramasinghe et al. 1974). Sadakane et al. (1985) found that the rotational broadening shown in the IUE spectrum of HD 77581 was similar to that seen in the spectrum of $\kappa$ Ori, and they concluded that $v_{\rm rot} \sin i = 80 \pm 10$ km s$^{-1}$. However, from an echelle CCD spectrum of $\kappa$ Ori taken in our February 1989 run, it is clear that the lines of $\kappa$ Ori are less broadened than those of HD 77581. Since all estimates seem essentially consistent with the result of Zuiderwijk (1995), we have used his result for the simulations. For $\beta$, we used, following Rappaport & Joss (1983), a range 0.9–1.0. The parameters we use and the results are listed in Table 6.

A problem that arose in doing the Monte-Carlo simulations was that with the parameters listed above for a large fraction of the trials ($f_{\rm bad}$ in Table 6), the parameters are inconsistent with each other in the sense that the width of the eclipse can not be reproduced, not even for an inclination of 90° and a filling factor of unity at periastron. It is not clear how to treat these events. One might just reject them, but in that way the estimates will be based on distributions of observed quantities which no longer reflect the actual ones. For the case mentioned, one expects that only trials with low $\theta_{\rm ecl}$, low $v_{\rm rot} \sin i$ and/or low $K_{\rm opt}$ will not be rejected. It is not clear what the net effect on the determination of $M_{\rm X}$ is, since trials with low $\theta_{\rm ecl}$, $v_{\rm rot} \sin i$ or $K_{\rm opt}$ lead to high, hardly changed or low $M_{\rm X}$, respectively.

The discrepancy between the predicted and observed eclipse angle might be due to systematic errors in $K_{\rm opt}$, $v_{\rm rot} \sin i$ or $\theta_{\rm ecl}$.

**Table 6.** Parameters of the Vela X-1/HD 77581 system

| Parameter | Value[a] | | | Unit |
|---|---|---|---|---|
| *Observed* | | | | |
| $T_{\pi/2}$[b] | 2444279.0466(37) | | | JD |
| $P_{\rm orb}$ | 8.964416(49) | | | day |
| $\dot{P}_{\rm orb}/P_{\rm orb}$ | $<1.9\,10^{-5}$ | | | yr$^{-1}$ |
| $a_{\rm X}\sin i$ | 112.98(35) | | | lt-s |
| $e$ | 0.0885(25) | | | |
| $\varpi_{\rm X}$ | 150.6(18) | | | ° |
| $\dot{\varpi}_{\rm X}$ | $<1.9$ | | | ° yr$^{-1}$ |
| $\theta_{\rm ecl}$ | 30–36 | | | ° |
| $v_{\rm rot}\sin i$ | 116(6) | | | km s$^{-1}$ |
| $K_{\rm opt}$ | 17.0–29.7 | 20.8(17) | 24.6(23) | km s$^{-1}$ |
| *Inferred* | | | | |
| $f_{\rm co}$ | $0.69^{+0.09}_{-0.08}$ | $0.69^{+0.09}_{-0.08}$ | $0.70^{+0.08}_{-0.08}$ | |
| $i$ | $>74$ | $>73$ | $>75$ | ° |
| $a$ | $53.4^{+1.6}_{-1.4}$ | $53.1^{+1.8}_{-1.1}$ | $53.6^{+1.4}_{-1.1}$ | $R_\odot$ |
| $R_{\rm opt}$ | $30.0^{+1.8}_{-1.9}$ | $30.2^{+1.7}_{-2.2}$ | $29.9^{+1.4}_{-1.8}$ | $R_\odot$ |
| $M_{\rm opt}$ | $23.5^{+2.2}_{-1.5}$ | $23.2^{+2.5}_{-1.1}$ | $23.6^{+1.9}_{-1.2}$ | $M_\odot$ |
| $M_{\rm X}$ | $1.88^{+0.69}_{-0.47}$ | $1.75^{+0.34}_{-0.33}$ | $2.08^{+0.46}_{-0.43}$ | $M_\odot$ |
| $f_{\rm bad}$[c] | 44 | 35 | 50 | % |

[a] For the observed quantities, errors are $1\sigma$ confidence limits, limits are 95% confidence, and ranges are 99% confidence. For the inferred quantities, the uncertainties and limits represent approximate 95% confidence limits. The three different sets of inferred parameters reflect the three distributions for $K_{\rm opt}$. For details and references, see Sect. 5
[b] Time of mean longitude 90°
[c] Fraction of trials rejected because eclipse width could not be fit; see Sect. 5

For $\theta_{\rm ecl}$, e.g., this could be due to partial eclipses by the stellar wind. However, most likely the cause is also at least partly to be found in the assumption that the star is in quasi-hydrostatic equilibrium with respect to the instantaneous potential. Noteworthy in this respect is that in the optical light curve of the system the two minima do not exactly coincide in time with superior and inferior conjunction, as would be expected if the star did adjust instantaneously to the varying potential (Tjemkes et al. 1986; Wilson & Terrell 1994). The minimum closest to X-ray eclipse is shifted by about 0.05 in orbital phase. Hence, at the time of X-ray eclipse the observed size of the star is not yet as small as one would expect, and the measured X-ray eclipse will be longer than the expected one. Furthermore, we notice that in the above it is assumed that the orbital and equatorial planes coincide. If supernova explosions are asymmetric, however, as is indicated by the high velocities found for single radio pulsars (e.g., Lyne & Lorimer 1994), this is not necessarily the case, although one might expect the deviation to be small in view of the relatively small eccentricity of the system.

We conclude that although it seems likely that the inclination will be close to 90°, we can not exclude lower values. For the lower limits to the mass of the neutron star, which is the most interesting quantity with respect to observational constraints on the equation of state, we will make the conservative assumption that $i=90°$ and we will use the limits on $K_{\rm opt}$ derived from the fit to all data. We find that with 95% confidence $M_{\rm X}>1.43\,M_\odot$ and with 99% confidence $M_{\rm X}>1.32\,M_\odot$.

## 6. Summary and conclusions

We have obtained a set of new high-quality spectra of HD 77581, the optical counterpart of Vela X-1. From these observations, we found that this star shows large excursions in velocity with respect to the expected radial-velocity curve. The changes are autocorrelated within one night, but not from one night to the other. The line profiles show strong, erratic changes, which are probably related to the velocity excursions. The profile changes are very similar for lines of different ions, indicating that they reflect large-scale motion of the surface rather than density enhancements in the wind. A possible cause could be that they are due to pulsations induced by the varying tidal force exerted by the neutron star in its eccentric orbit.

Due to the presence of the erratic excursions in radial velocity, the high accuracy of the velocity determinations became of little value for the mass determinations. Therefore, we also made radial-velocity determinations for an older set of digitized photographic spectra, as well as for the available IUE spectra of the star. We analysed all velocities using a Monte-Carlo method and determined that the uncertainty induced by the velocity excursions was about 7%. We argued that although we do not find evidence for effects related to the tidal deformation, these effects may well be present at the $\lesssim 5\%$ level.

The largest uncertainty is due to the possible presence of a systematic positive deviation of the radial velocities around the time of velocity minimum. Our simulations showed that the effect was significant at the $>95\%$ confidence level. Lacking a physical understanding of the cause of the effect, however, we believe that for a conservative lower limit on the mass of the neutron star, one should use the limit derived using $K_{\rm opt}$ found from the fit to all data, i.e., $1.43\,M_\odot$ (95% confidence). This value does not allow one to strongly constrain the neutron-star equation of state (for a discussion of all neutron-star mass determinations from X-ray pulsars, see Van Kerkwijk et al. 1995). We notice, though, that if the effect were real, and no systematic deviations occurred at other orbital phases, the 95% confidence lower limit to the mass of Vela X-1 would be $1.74\,M_\odot$.

Given our results, it follows that in order to derive more accurate constraints on the orbital parameters – and thus on the mass of this possibly very massive neutron star – one needs to understand the tidal interaction between the neutron star and its companion. For this purpose, it is probably necessary to study both the short and long-term behaviour of the system. For the latter, it is most important to obtain an idea of the statistical behaviour of the system, and since the velocity shifts one wants to measure are $\sim 5\,{\rm km\,s^{-1}}$, very high accuracy is not necessary. Also, since the deviations are correlated within single nights, only one spectrum needs to be taken each night. For the short-time study, however, high resolution observations covering the whole orbit are probably essential, in order to fol-low the pulsations and interaction in detail. It may be that even with great effort, it will not be possible to obtain a determination of the neutron-star mass that does not depend strongly on the assumed model of the structure of the star and the tidal interaction. However, independent of whether this is the case, the system provides for a rather clean test site of the effects of tidal interaction on a star, since the neutron star is a point mass for all practical purposes.

*Acknowledgements.* We thank Jochen Eislöffel and Werner Verschueren for useful discussions about the reduction of echelle spectra and the cross-correlation of early-type spectra, and Thomas Augusteijn, Alan Levine and Saul Rappaport for help with the Monte Carlo method for the mass determinations. EJZ acknowledges use of the computing facilities at SARA (Amsterdam) and CERN (Geneva), MHvK support by NASA through a Hubble fellowship (HF-1053.01-93A) awarded by the Space Telescope Science Institute, and CS a research grant from the Belgian Fund for Scientific Research (NFWO).